\documentclass[12pt]{iopart}
\usepackage{graphicx}
\usepackage{iopams}  
\usepackage[colorlinks=true,citecolor=blue]{hyperref}
\begin{document}

\title{Universal three-body physics at finite energy near Feshbach resonances}

\author{Yujun Wang\footnotemark and B. D. Esry}
\address{Department of Physics, Kansas State University, Manhattan, Kansas, 66506, USA}
\footnotetext{Present address: JILA, University of Colorado, 440 UCB, Boulder, Colorado, 80309, USA}

\begin{abstract} 
We find that universal three-body physics extends beyond the threshold regime to non-zero energies.
For ultracold atomic gases with a negative two-body $s$-wave scattering length near a Feshbach resonance,
we show the resonant peaks characteristic of Efimov physics persist in three-body recombination to higher 
collision energies.  For this and other inelastic processes, we use the adiabatic hyperspherical representation 
to derive universal analytical expressions for their dependence on the scattering length, the collision
energy, and --- for narrow resonances --- the effective range.  These expressions are 
supported by full numerical solutions of the Schr\"odinger equation and display log-periodic
dependence on energy characteristic of Efimov physics.  This dependence is robust and might be used to 
experimentally observe several Efimov features.
\end{abstract}
%\pacs{}
\maketitle

\section{Introduction}
Ultracold quantum degenerate gases near Feshbach resonances have been studied intensively in recent years~\cite{FeshbachChin}. 
The tunability of the interactions between the atoms, characterized by their $s$-wave scattering length $a$, 
makes ultracold quantum gases an ideal testbed to study a variety of few-body problems.
Three-body systems in the strongly interacting regime $|a|\rightarrow\infty$ especially interest people due to their importance in many areas in physics,
including condensed matter, atomic, molecular and nuclear physics~\cite{Jensen2004,Braaten2006}. 

Central among all ultracold three-body phenomena is the Efimov effect, which occurs 
when the scattering length $a$ is much larger than the characteristic length scale $r_0$ of the atomic interaction~\cite{Esry1999,Nielsen1999}. 
First predicted in the early 70's by the nuclear physicist Vitaly Efimov~\cite{Efimov1970}, 
this effect refers to the existence of an infinite number of weakly bound three-body states when $|a|\rightarrow\infty$.
It has a profound impact on the three-body scattering observables and
leads to universal three-body behavior irrespective of the short-range interactions between the particles~\cite{Braaten2006}.
In particular, when $a<0$, the rate $K_3$ for three-boson recombination, $B+B+B\rightarrow B_2+B$, features
a log-periodic series of resonance peaks on top of an overall $a^4$ scaling near zero energy. Explicitly~\cite{Braaten2006,Esry1999,Nielsen1999},
in atomic units,
\begin{eqnarray}
K_3\!=\!\frac{4590\sinh(2\eta)}{\sin^2[s_0\ln(|a|/r_0)\!+\!\Phi\!+\!1.53]\!+\!\sinh^2\eta} \frac{a^4}{m},
\label{eqn_K3_zero}
\end{eqnarray}
where $s_0\approx 1.00624$ is a universal constant, $\Phi$ is the three-body short-range phase~\cite{Braaten2006,ScalingDIncao2005,PhaseDIncao}, $\eta$ is a parameter characterizing 
short-range inelastic transitions~\cite{Braaten2006,ScalingDIncao2005}, and $m$ 
is the atomic mass.  Physically, 
the peaks appear in $K_3$ when an Efimov state moves across the three-body break-up threshold and
becomes bound~\cite{Esry1999,Esry2007}. Another three-body inelastic process which is also important in ultracold 
experiments is the vibrational relaxation of a weakly-bound molecule: 
$B_2^*+B\rightarrow B_2+B$ ($a>0$). The rate for vibrational relaxation $V_{\rm rel}$ has the same functional form as Eq.~(\ref{eqn_K3_zero}), 
with the overall $a^4$ scaling changed to just $a$; the overall coefficient, to 20.03; and the constant in the phase, to 1.47. 
 
Near a narrow Feshbach resonance, however, the universal three-body physics is 
modified~\cite{Petrov2004,Gogolin2008,WangPrepNarrow}. 
The dependence of the three-body physics on the resonance width in magnetic field can be parametrized with the two-body effective range $r_{\mathrm{eff}}$, which is inversely proportional to the resonance width~\cite{Petrov2004}. 
It has recently been predicted that near a narrow resonance three-body inelastic processes dominated by short-range transitions are 
suppressed for bosons and enhanced for mixed spin fermions~\cite{WangPrepNarrow}.

The Efimov and universal three-body physics introduced above are studied through the dependence of the three-body observables on $a$ in the limit of
zero scattering energy.  Studies away from zero energy have mainly considered only 
the corrections to the zero-energy results~\cite{ThermDIncao2004,Jonsell2006,Braaten2007,Platter2008,Massignan2008,Braaten2008}.
But, interestingly, it has recently been shown that the energy dependence also displays
signatures of the Efimov effect for $a>0$~\cite{Wang2010}.  For energies
beyond the threshold regime, however, the higher partial wave contributions to the inelastic rates are not suppressed by the Wigner threshold law~\cite{Esry2001} 
and tend to mask these energy-dependent Efimov features.  These contributions, combined with thermal averaging effects, make
observing energy-dependent Efimov features in this case quite challenging.  To overcome these difficulties, a scheme to collide
Bose-Einstein condensates has therefore been proposed~\cite{Wang2010}.

In this paper, we show that signatures of the Efimov effect also appear at non-zero collision energies in recombination for $a<0$. 
In contrast to the finite-energy Efimov features identified previously~\cite{Wang2010}, 
the Efimov signatures we show here have negligible higher partial wave contributions and are much less
sensitive to thermal averaging. We also study the energy dependence of the vibrational relaxation rate when $a>0$.
Finally, we show that Efimov and universal features remain for large negative $r_{\rm eff}$, corresponding to a 
narrow Feshbach resonance.  In most cases, we find analytic expressions 
for the rates and confirm them with numerical solutions of the Schr\"odinger equation.

\section{Method}
\label{Method}
To study three-body universality at finite energies, we have numerically solved the three-body Schr{\"o}dinger equation for identical bosons using the adiabatic hyperspherical 
representation~\cite{HeCalc2}.
In this representation, the Schr{\"o}dinger equation reduces to a set of coupled equations for the hyperradial wavefunctions $F_\nu(R)$,
\begin{eqnarray}
\left[-\frac{1}{2\mu}\frac{d^2}{d R^2}+W_{\nu\nu}\right ]F_\nu+\sum_{\nu'\neq\nu}W_{\nu\nu'}F_{\nu'}=E F_\nu.
\label{Eq_Hyper}
\end{eqnarray}
The hyperradius $R$ is a measure of the overall size of the system and $\mu=m/\sqrt{3}$ is the three-body reduced mass. 
The effective three-body potentials are
\begin{eqnarray}
W_{\nu\nu}(R)=U_\nu(R)-\frac{1}{2\mu}Q_{\nu\nu}(R), 
\end{eqnarray}
where $U_\nu(R)$ are calculated from the adiabatic equation
\begin{eqnarray}
H_{\rm ad}(R; \Omega)\Phi_\nu(R;\Omega)=U_\nu(R)\Phi_\nu(R;\Omega).
\label{eqn_AdEqn}
\end{eqnarray}
The adiabatic Hamiltonian $H_{\rm ad}$ includes the hyperangular kinetic energy and all interactions.
The diagonal couplings $Q_{\nu\nu}(R)$ are determined from 
\begin{eqnarray}
Q_{\nu\nu}=\left<\!\!\left<\frac{d\Phi_\nu}{dR} \biggl| \frac{d\Phi_\nu}{dR}\right>\!\!\right>,
\end{eqnarray}
where the double brackets indicate integration over the hyperangles $\Omega$.
The non-adiabatic couplings $W_{\nu\nu'}$ ($\nu\neq \nu'$) govern the transitions between channels.
The effective three-body potentials $W_{\nu\nu}$ are crucial for the study of three-body scattering, especially at low energies 
since they determine the universal behavior of the rates like Eq.~(\ref{eqn_K3_zero})~\cite{ScalingDIncao2005}.

To solve Eq.~(\ref{eqn_AdEqn}) numerically, we must specify the interaction potential.  Fortunately,
when $|a|\gg r_0$, the universality of low-energy three-body physics allows us to model the interactions between the 
atoms near a Feshbach resonance by single channel interactions with large $|a|$, 
as the details of the atomic interactions affect three-body observables only through the short-range parameters $\Phi$ and $\eta$. 
This model is equivalent to a multi-channel treatment as long as there is only one open channel in the two-body system.
Therefore, when solving Eq.~(\ref{eqn_AdEqn}), we simply use a pair-wise sum of two-body potentials 
\begin{eqnarray}
v(r)=-D\mathrm{sech}^2(r/r_0),
\label{eqn_potential}
\end{eqnarray}
where $r$ is the interparticle distance and $D$ is a parameter that we use to control the scattering length as well as the number of two-body bound states.

Our numerical approach has been described in detail in Refs.~\cite{HeCalc2,EsryThesis,Aymar1996,BurkeThesis}.  Briefly, 
we first solve Eq.~(\ref{eqn_AdEqn}) by expanding $\Phi_\nu(R;\Omega)$  on Wigner $D$ functions for the Euler angle
dependence of the overall rotations~\cite{HeCalc2} and on 
a two-dimensional, direct product  B-spline basis for the hyperangles representing relative motion in the body frame~\cite{EsryThesis}. 
Inelastic rates are then calculated by solving 
Eq.~(\ref{Eq_Hyper}) with the eigenchannel $R$-matrix method~\cite{Aymar1996,BurkeThesis}.

\begin{figure}
\begin{center}
\includegraphics[clip=true,scale=0.9]{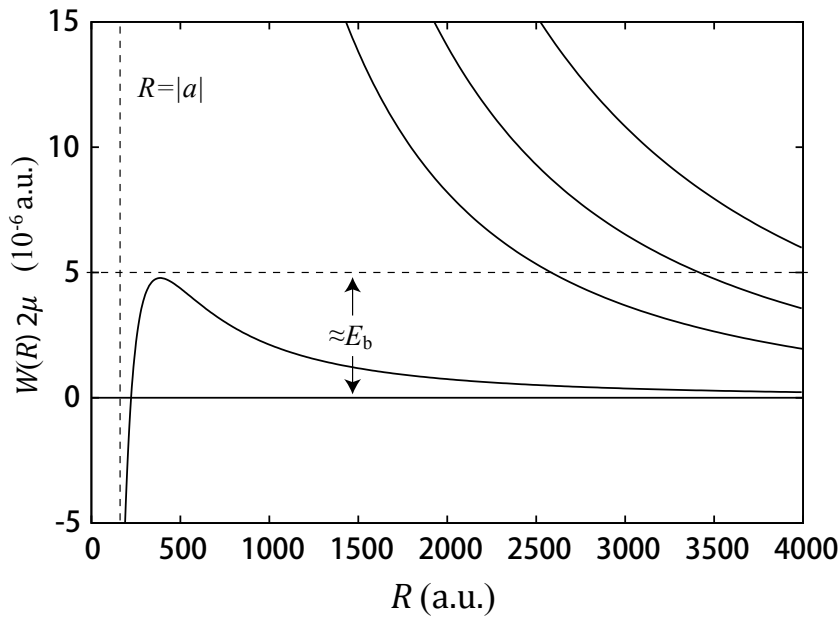}
\end{center}
\caption{The adiabatic hyperspherical potentials for $a$=--180~a.u. 
The deeply bound atom-molecule potentials are not shown, so all of the curves represent the three-body 
continuum, $B$+$B$+$B$, asymptotically. 
}
\label{Fig:Pot} 
\end{figure}

We use these numerical solutions to verify our analytical expressions for the universal behavior.  These expressions are obtained by 
solving the hyperradial equation Eq.~(\ref{Eq_Hyper}) using the universal results for the hyperradial potential $W_{\nu\nu}(R)$~\cite{Efimov1973,ScalingDIncao2005}.
Moreover, for recombination with $a<0$ and vibrational relaxation with $a>0$, we need only know the behavior of the initial channel.
In Fig.~\ref{Fig:Pot}, we show $W_{\nu\nu}(R)$ for $a$=--180~a.u. numerically 
calculated from Eq.~(\ref{eqn_AdEqn}).  
The lowest energy curve is the initial channel for recombination.  It has a barrier at $R\approx |a|$~\cite{EsryJPBLetter}. %BDE
For smaller $R$, the
potential has Efimov character; and for larger $R$, free particle character.
For $a<0$ and $|a|\gg r_0$, we can idealize the behavior in the incident channel for recombination as
\begin{eqnarray}
W_{\nu\nu}(R)=\left\{
\begin{array}{cc}
\displaystyle -\frac{s_0^2+1/4}{2\mu R^2} &  r_0\ll R\ll |a|,\\
\displaystyle \frac{15/4}{2\mu R^2} & R\gg |a|.
\end{array}\right.,
\label{Eq:NarrowPotential}
\end{eqnarray}
For $R\le r_0$, $W_{\nu\nu}$ is not universal and the solution from that region will simply be parametrized at $R=r_0$
with $\Phi$ and $\eta$. 
For vibrational relaxation when $a>0$, the potential for the incident channel is only changed for $R\gg a$ to
\begin{eqnarray}
W_{\nu\nu}=-E_{2b}+\frac{l(l+1)}{2\mu R^2},
\end{eqnarray}
where $E_{2b}=1/m a^2$ is the binding energy of the weakly-bound molecule and
$l$ is the angular momentum between the atom and the molecule.

For both processes, the coupling between the initial channel and the final deeply-bound atom-molecule channel is only significant for $R\lesssim r_0$, 
and we can treat the channels as uncoupled for $R>r_0$.  The solutions are thus completely determined by the potential Eq.~(\ref{Eq:NarrowPotential}),
and, by assuming each potential Eq.~(\ref{Eq:NarrowPotential}) holds over the entire domain,
the hyperradial wave functions $F_\nu(R)$ are~\cite{Jonsell2006}
\begin{eqnarray}
F_\nu(R)\!=\!\left\{
\begin{array}{lc}
C_0\sin(\Phi+i\eta) & R= r_0,\\
C_1R^{1/2}\{J_{i s_0}(kR)\!-\!\tan\delta_1N_{i s_0}(kR)\} & \!\!\!\!\!\!\!\!r_0\le R\le \beta|a|,\\
C_2R^{1/2}\{J_{l_{\rm eff}\!+\!1/2}(kR)\!-\!\tan\delta N_{l_{\rm eff}\!+\!1/2}(kR)\} & R\ge \beta|a|, 
\label{Eq:AsymSoln}
\end{array}\right.
\end{eqnarray}
where $J_\nu(x)$ and $N_\nu(x)$ are Bessel functions of the first and second kind, respectively; and $k=\sqrt{2\mu E}$ with $E$ the
total energy relative to the three-body break-up threshold.  Since the boundary between
regions is not exactly at $R=|a|$ and cannot be precisely defined, we include the constant $\beta$ as a free parameter to be determined by
matching to numerical solutions~\cite{WangPrepNarrow}. 
The effective angular momentum $l_{\rm eff}$ is $3/2$ for recombination~\cite{Esry2001} and $0$ for relaxation.
The parameters $\Phi$ and $\eta$ are introduced in the same fashion as in Refs.~\cite{Braaten2006,ScalingDIncao2005} to
characterize the non-universal, short-range physics, including the inelastic transition.  They will generally be taken to
be free parameters and fit to the rates for a particular system.

By matching the hyperradial wavefunction $F_\nu(R)$ at each boundary, the asymptotic phase shift $\delta$ can be expressed
in terms of $\beta|a|$ and the short-range parameters at any energy.  To obtain the probability for inelastic transitions in such a single-channel model, 
we first calculate the probability for elastic scattering ${\cal R}$ which is given by the reflection coefficient:
\begin{eqnarray}
{\cal R}=\left|\frac{1-i\tan\delta}{1+i\tan\delta} \right|^2.
\end{eqnarray}
So long as $\eta$ is not zero (which corresponds to no coupling to deeply-bound states), $\delta$ will be complex and $\cal R$ will be less than unity.  
The overall probability for an inelastic transition is thus ${\cal P}=1-{\cal R}$.
For three-body recombination of identical bosons, the recombination rate $K_3$ is related to $\cal{P}$ by~\cite{Esry1999} 
\begin{eqnarray}
K_3=\frac{192\sqrt{3}\pi^2}{mk^4}{\cal P}.
\label{Eq_K3Def}
\end{eqnarray}
Equation~(\ref{eqn_K3_zero}) can then be reproduced by deriving $\cal P$ for recombination near zero energy ($k\rightarrow 0$).

\section{Three-body inelastic processes near broad Feshbach resonances}

\subsection{Three-body recombination at finite energies}
 
In general, each region identified in Eq.~(\ref{Eq:NarrowPotential}) becomes important when the de Broglie wavelength becomes
small enough to sample that length scale.
Specifically, at energies above $E_{\rm s}=1/\mu r_0^2$, the details in the short-range region will start to play a role and the rates will not be universal. 
Below $E_{\rm s}$ --- that is, at $R\geq r_0$ --- the three-body potentials and couplings are all universal. We therefore expect universal behavior 
in the three-body scattering for energies below $E_{\rm s}$.

Because the Efimov region of the lowest potential 
in Fig.~\ref{Fig:Pot} and Eq.~(\ref{Eq:NarrowPotential}) has a barrier at $R\approx |a|$, 
it can support a three-body shape 
resonance when an Efimov state exists behind the barrier.
When the energy of the Efimov state is above the three-body break-up threshold $E$=0, 
increasing the collision energy across the resonance energy will 
lead to a resonant peak in $K_3$~\cite{Esry1999,Esry2007}. 
This can be understood from the fact that near the resonant energy, the amplitude of the hyperradial wavefunction behind the barrier in $W_{\nu\nu}$ is greatly enhanced.
The three-body system thus has much larger probability in the small-$R$ region where the coupling between the incident channel and the deeply-bound atom-molecule channel 
peaks.
However, because the height of the barrier is $\sim E_{\rm b}=1/\mu a^2$, the resonant behavior in $K_3$ can only
be observed for energies below $E_{\rm b}$. And, below this energy, the barrier can only confine a 
resonant Efimov state with size on the order of $|a|$,
implying that at most one Efimov resonance can be observed as a function of energy for a fixed $a$.

\begin{figure}
\begin{center}
\includegraphics[clip=true,scale=0.9]{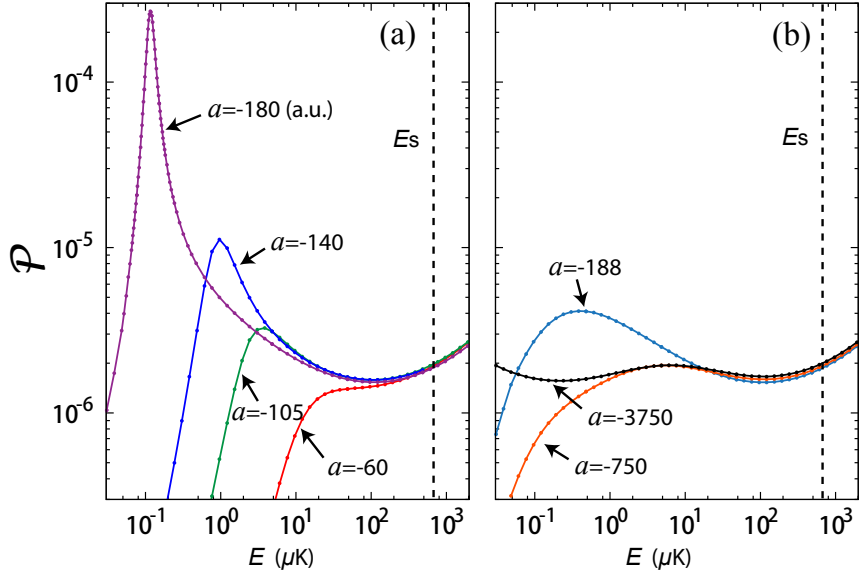}
\end{center}
\caption{The three-body recombination probability ${\cal P} (a<0)$ for three identical atoms 
with the mass of Cs when a resonant Efimov trimer state is 
above the three-body break-up threshold~\cite{Unit}. 
(a) As $|a|$ increases, the resonant peak moves towards the break-up threshold.
(b) Increasing $|a|$ further after the Efimov state becomes bound, the resonant peak disappears, 
leaving an oscillatory structure behind.} 
\label{fig_recomb_broad_resonance}
\end{figure}

In Fig.~\ref{fig_recomb_broad_resonance}, we show the evolution of the first Efimov resonance peak 
from finite energy to zero energy as $|a|$ increases (see also Ref.~\cite{Esry2007}). 
These recombination probabilities were calculated numerically.
Interestingly, when a peak moves below zero energy, an oscillatory structure with one full period appears
in $\cal P$ for $E>E_{\rm b}$ as shown in Fig.~\ref{fig_recomb_broad_resonance}(b). 

\begin{figure}[h]
\begin{center}
\includegraphics[height=2.5in]{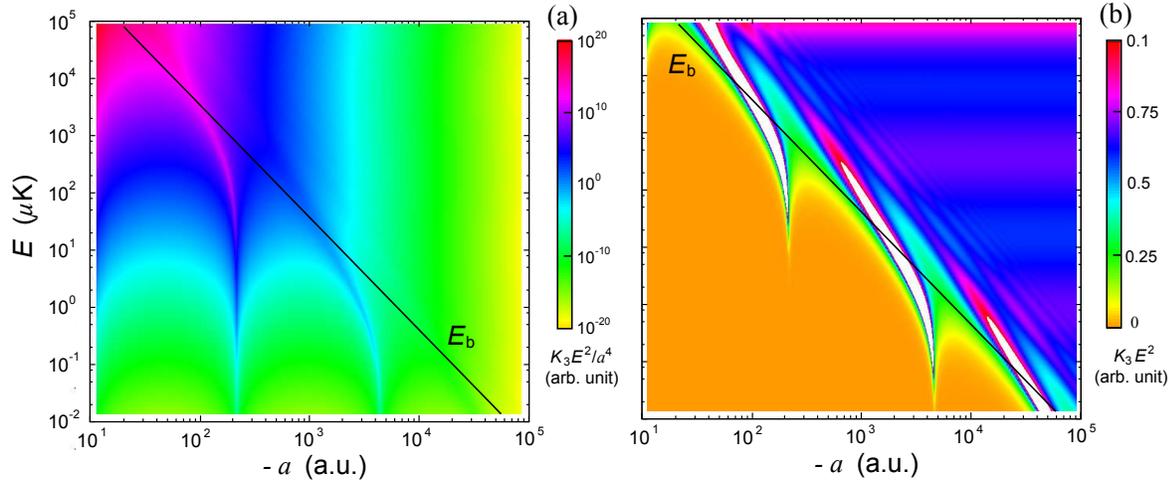}
\end{center}
\caption{The universal dependence of $K_3$ on scattering length and energy for identical bosons with the mass of $^7$Li %BDE
with $J$=0 as evaluated from Eqs.~(\ref{Eq_K3Def}) and (\ref{Eq:FiniteEnergy}). (a) The rate scaled as $K_3E^2/a^4$
to show better the resonant peaks. (b) The same data scaled as $K_3E^2$ to emphasize the energy-dependent oscillation 
described by Eq.~(\ref{eqn_prob_broad}) which can be seen in the upper right half of the plot.
Note that the fine structure parallel to the $E_{\rm b}$ line is an artifact of the discontinuity of our 
idealized potential in Eq.~(\ref{Eq:NarrowPotential}) at $R=\beta|a|$.
In both plots, $\Phi$=$\pi/2$, $\eta$=$0.02$ and $r_0$=15 a.u..
}
\label{Fig:K3:aN:En}
\end{figure}

To understand the oscillations revealed in these numerical results, 
we employ the analytical model introduced in Sec.~\ref{Method}. Matching the hyperradial wavefunctions at finite energies
gives
\begin{eqnarray} 
\fl {\cal R}=\left|\left(\frac{(J_2'\!-\!i N_2')[J_{i s_0}\!-\!\tan\delta_1N_{i s_0}]}
{(J_2\!-\!i N_2)[J_{i s_0} '\!-\!\tan\delta_1 N_{i s_0} ']}-1\right)\!\! 
\left(\frac{(J_2'+i N_2')[J_{i s_0}\!-\!\tan\delta_1N_{i s_0}]}
             {(J_2+i N_2)[J_{i s_0}'\!-\!\tan\delta_1N_{i s_0} ']}-1\right)^{-1}\right|^2.
\label{Eq:FiniteEnergy}
\end{eqnarray}
The primes denote derivatives, and the Bessel
functions should be evaluated at $R=\beta|a|$.
When the scattering energy satisfies $k\ll 1/r_0$, $\tan\delta_1$ is given by
\begin{eqnarray} 
\fl
\tan\delta_1\approx\frac{\sin(\phi+i\eta-i\pi s_0/2)}{\cos(\phi+i\eta+i \pi s_0/2)},
\qquad
\tan\phi=\tan[\Phi-s_0\ln(k r_0)+\varphi_0+\pi/4],
\label{Eq:PhaseShift}
\end{eqnarray}
with
\begin{eqnarray}
\tan\varphi_0=\frac{\mathrm{Re}[\Gamma(i s_0)]-\mathrm{Im}[\Gamma(i s_0)]}{\mathrm{Re}[\Gamma(i s_0)]+\mathrm{Im}[\Gamma(i s_0)]}.
\end{eqnarray}
For identical bosons, $\varphi_0$=$-0.15418\pi$. 
Equations~(\ref{Eq:FiniteEnergy}) and (\ref{Eq:PhaseShift}) give a complete universal expression for $K_3$ in terms of the scattering length and energy.
This analytic expression is evaluated and plotted in Fig.~\ref{Fig:K3:aN:En}. In (a), the evolution of the resonant peaks with energy can be clearly seen, but the oscillatory structure 
--- which should appear at fixed, large $|a|$ as a function of $E$ ---
is essentially invisible due to its relatively small modulation compared to the resonant peaks.  Figure~\ref{Fig:K3:aN:En}(b) replots the data to better
show the energy-dependent oscillations.  
Note that the fine structure parallel to the $E_{\rm b}$ line is an artifact of the discontinuity of our
idealized potential in Eq.~(\ref{Eq:NarrowPotential}) at $R=\beta|a|$.
The connection between the Efimov resonances and the energy-dependent 
oscillations is evident in Fig.~\ref{Fig:K3:aN:En}(b). The peak of an oscillation 
appears approximately at the energy where 
the zero-energy position of each Efimov resonance  intersects the $E_{\rm b}$ line. The rate oscillates through a full period when an Efimov 
resonance moves below the 
three-body breakup threshold.  Therefore, the number of bound Efimov states can be directly read from the number of full oscillations.

Remarkably, when $\eta\ll 1$ (corresponding to a small probability for an inelastic transition at small $R$ as is expected), 
Eqs.~(\ref{Eq:FiniteEnergy}) and (\ref{Eq:PhaseShift}) reduce to a simple form 
in the energy range $E_{\rm b}\ll E\ll E_{\rm s}$ ($kr_0\ll 1$):
\begin{eqnarray}
{\cal P}=\frac{2\sinh(\pi s_0)\sinh(2\eta)}{\cosh(\pi s_0)+\sin[-2s_0\ln(kr_0)+2\Phi-2\varphi_0]},
\label{eqn_prob_broad}
\end{eqnarray}
It should be noted that for $E>E_{\rm b}$ the parameters $\Phi$ and $\eta$ will depend on energy
in general. However, for energies $E\ll E_{\rm s}$ the change in the shape of the short-range hyperradial wavefunction is small since $E_{\rm s}$ is still
small compared to the depth of the potential,
therefore both $\Phi$ and $\eta$ can be treated as energy-independent quantities.

The log-periodicity of the energy-dependent oscillations in Eq.~(\ref{eqn_prob_broad}) with period $e^{2\pi/s_0}$ provides the clear connection
with Efimov physics.  Like the energy-dependent oscillations observed for $a>0$~\cite{Wang2010},
the overall phase of the oscillations does not depend on $a$ as is clearly shown in Fig.~\ref{Fig:K3:aN:En}(b). 
Therefore, as $|a|$ increases, the lower limit of validity $E_{\rm b}=1/\mu a^2$ 
of Eq.~(\ref{eqn_prob_broad}) decreases and 
more oscillations will appear towards the zero-energy limit.  At the same time, those that
have already appeared at higher energies remain unchanged. 

These oscillations might thus provide one experimentally accessible means of measuring multiple Efimov features.
In the experiments observing the Efimov effect via zero-energy three-body 
recombination~\cite{Kraemer2006,Zaccanti2009,Pollack2009,Ottenstein2008,Huckans2009,Barontini2009,Gross2009,Williams2009}, 
the temperature must be kept below $E_{\rm b}$ as $|a|$ is increased~\cite{ThermDIncao2004}.
This restriction makes the observation of multiple Efimov features quite challenging.
However, if the Efimov features are measured via the energy-dependent oscillations, the experimental requirements might be less demanding.
Furthermore, the $a$-independence of Eq.~(\ref{eqn_prob_broad}) allows the energy-dependent oscillations to be measured 
at different fixed $a$, provided that $E\geq E_{\rm b}$ ($k|a|\geq 1$).
Because the modulation of the oscillations is small compared to the resonant peaks, though, 
the scattering length where the measurements are carried out should be chosen to be far away from resonant peaks.

\begin{figure}
\begin{center}
\includegraphics[clip=true,scale=0.9]{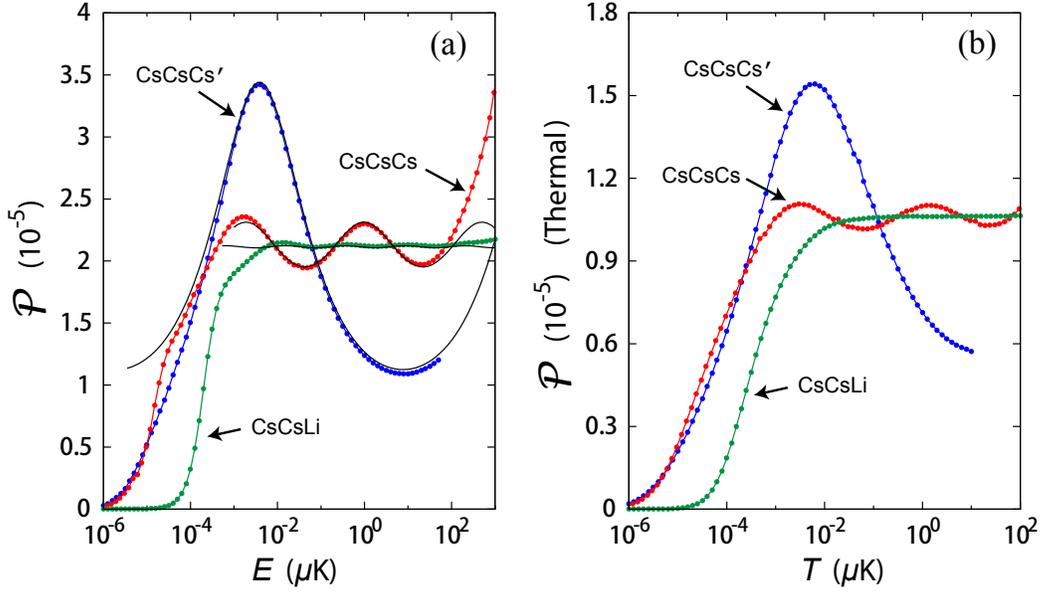}
\end{center}
\caption{(a) The three-body recombination probability ${\cal P}(E)$ for three identical Cs atoms, 
two identical Cs atoms with one Cs atom in a 
different spin state (Cs$'$) and two identical Cs atoms with one Li atom. 
The symbols are numerical results, and the black solid lines are the analytical result from Eq.~(\ref{eqn_prob_broad}). 
The analytical curves are only shown over their range of validity, $E_{\rm b}\leq E \leq E_{\rm s}$,
but $E_{\rm s}$ is off the scale.
For CsCsCs, $a=2\times 10^5$~a.u.; for CsCsCs$'$, $a=6\times 10^5$~a.u.; and 
for CsCsLi, $a=10^5$~a.u. 
(b) The recombination probability after thermal averaging. 
To make clear the comparison of the relative modulation in the oscillations from the three different 
systems, ${\cal P}$ for CsCsCs$'$ and CsCsLi has been multiplied by 5 and 1.9, respectively.
}
\label{fig_recomb_broad_oscillations}
\end{figure}

Our results apply equally well to three-body systems $BBX$ with two identical bosons, 
assuming $a\gg r_0$ is the $B$+$X$ scattering length.
All that changes is the value of $s_0$~\cite{MassDIncao2006}. 
The density of Efimov states in such systems grows with the mass ratio $\gamma=m_B/m_X$
since $s_0$ grows with $\gamma$, decreasing the spacing $e^{2 \pi/s_0}$ of energy-dependent Efimov 
features~\cite{MassDIncao2006,Efimov1973}.  
As can be seen from Eq.~(\ref{eqn_prob_broad}), however, the relative size of the modulation of $\cal P$ is 
\begin{eqnarray}
\Delta {\cal P}=2\frac{\cosh(\pi s_0)}{\sinh^2(\pi s_0)}
\label{Eq:RelMod}
\end{eqnarray}
which increases with decreasing $s_0$. 
The two
effects thus conflict: to see more Efimov features requires a large $s_0$ ($\gamma$), but seeing them clearly requires a small $s_0$ ($\gamma$).
Figure~\ref{fig_recomb_broad_oscillations}(a) shows ${\cal P}\propto E^2 K_3$ [see Eq.~(\ref{Eq_K3Def})] for model systems chosen to have mass ratios matching representative
atomic systems. For three identical Cs atoms, Eq.~(\ref{Eq:RelMod}) predicts $\Delta{\cal P}$ 
to be 0.170; for two identical Cs atoms and one Cs atom in a different spin state, the ratio is 1.37; 
and for two identical Cs atoms with one Li atom, the ratio is only 0.00789.  The predicted spacing between features is 515,
3.9$\times$10$^6$, and 23.8, respectively.  The analytic expression, Eq.~(\ref{eqn_prob_broad}), was fit to the numerical data by adjusting
only $\Phi$ and $\eta$ and is also plotted in Fig.~\ref{fig_recomb_broad_oscillations}.  It agrees quite well
 with the numerical results. 

A practical balance of contrast versus spacing can likely be found by first determining the 
smallest contrast experimentally observable.  
This choice makes $s_0$ as large as experimentally tolerable and thus 
gives the smallest spacing between features.  For instance, if a relative modulation of 
$\Delta{\cal P}$=10\% can be observed, then from Eq.~(\ref{Eq:RelMod}) we find $s_0=1.1748$. 
The corresponding mass ratio is $\gamma\approx 7.1$,
and the spacing between Efimov features is 210 --- a considerable improvement over three identical bosons.  

To determine whether these features are experimentally observable even in the presence of a distribution 
of collision energies, we must thermally average the energy-dependent rates. 
Assuming a Boltzmann distribution~\cite{Suno2003,ThermDIncao2004}, we find the results 
shown in Fig.~\ref{fig_recomb_broad_oscillations}(b).
For $E_{\rm b}\leq E\leq E_{\rm s}$, the oscillatory structure in ${\cal P}$ generally becomes less clear after thermal averaging.
For $BBX$ three-body systems  where $s_0$ is small, such as CsCsCs$'$,  
the oscillations are well preserved after thermal averaging, with the trade-off that the oscillation periods are huge.

\subsection{Vibrational relaxation at finite energies}

Although the relaxation rates $V_{\mathrm{rel}}$ for three identical bosons with $a>0$ 
resemble the $a<0$ recombination rates in the zero-energy limit~\cite{Braaten2006,ScalingDIncao2005},
they differ dramatically at higher energies. Note that zero scattering energy now 
refers to the atom-molecule breakup threshold rather than 
the three-body breakup threshold as for recombination.
The usual two-body Wigner threshold law gives a constant rate for atom-molecule collisions 
in the threshold regime $E<E_{\mathrm{th}}$ where generally $E_{\mathrm{th}}\approx E_{\rm b}$. 
An exception arises when the zero energy value 
of $V_{\mathrm{rel}}$ is near a resonant peak 
which occurs when the atom-molecule scattering length $a_{\rm am}$ is much larger than $r_0$.  
In this case $E_{\mathrm{th}}\approx 1/\mu_{\rm am} a_{\mathrm{am}}^2$, where $\mu_{\rm am}$ 
is the atom-molecule reduced mass. 
For $E_{\rm th}\leq E \leq E_{\rm s}$, $V_{\mathrm{rel}}$ scales like 
\begin{eqnarray}
V_{\rm rel}=C k^{-1}. 
\label{Eq:BroadVrel}
\end{eqnarray}
The constant $C$ depends on the short-range physics and is related to the short-range parameter $\eta$. But different from three-body recombination,
the relaxation rates does not show any oscillations in our numerical calculations.

\section{Three-body inelastic processes near narrow Feshbach resonances}

Following Ref.~\cite{WangPrepNarrow} , we model a three-body system near a narrow Feshbach resonance by using pairwise two-body potentials 
that support an $s$-wave shape resonance.  We accomplish this by adding a barrier to our model two-body potentials, ensuring that the resonance has
a large, negative effective range $r_{\rm eff}$~\cite{Petrov2004,Gogolin2008}.
For such $r_{\rm eff}$, the potential $W_{\nu\nu}(R)$ in the region $r_0 \ll R\ll |r_{\mathrm{eff}}|$
is replaced by a weak, non-universal Coulomb potential~\cite{WangPrepNarrow}.
For identical bosons, when $|a|\gg |r_{\mathrm{eff}}|\gg r_0$ the new length scale 
produces a $1/|r_{\mathrm{eff}}|$ 
suppression in the zero-energy inelastic rates that lead to a deeply bound two-body state~\cite{WangPrepNarrow}.

This new length scale in $W_{\nu\nu}(R)$ introduces another energy scale for the universal behavior
when the wavelength is short enough to sample 
the three-body potentials and couplings in $r_0 \ll R\ll |r_{\mathrm{eff}}|$.
The new energy scale is $E_{\rm eff}=1/\mu |\alpha r_{\mathrm{eff}}|^2$ and defines a new
energy region: $E_{\rm eff} < E< E_{\rm s}$. The parameter $\alpha\approx 0.28$ 
plays a role similar to $\beta$ and 
was determined in Ref.~\cite{WangPrepNarrow} by fitting the universal formula for $K_3$ near 
a narrow Feshbach resonance to numerical results near zero-energy.
Although it was determined at zero energy, it should apply at non-zero energies as well.
Where we treated these processes at zero energy in Ref.~\cite{WangPrepNarrow}, we will now find the 
non-zero energy expressions for their rates. 

\subsection{Three-body recombination at finite energies}

\begin{figure}
\begin{center}
\includegraphics[clip=true,scale=0.9]{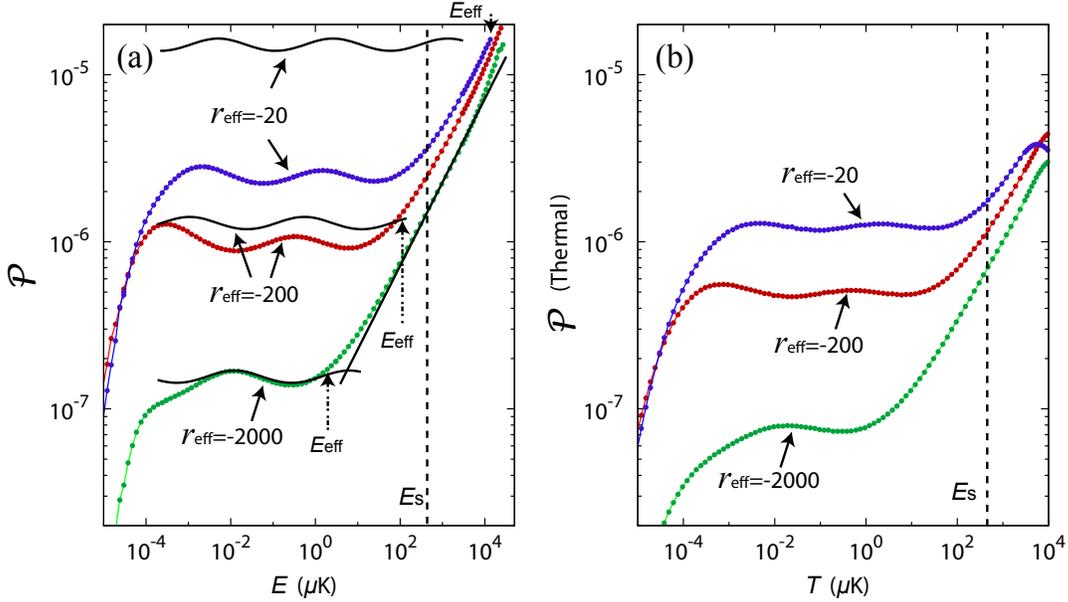} 
\end{center}
\caption{(a) The three-body recombination probability ${\cal P} (a<0)$ for identical bosons near narrow Feshbach resonances with $r_{\mathrm{eff}}$=$-$20, $-$200, and $-$2000~a.u.. 
For all cases, $a$=$-10^5$~a.u., and $r_0\approx 50$~a.u.. 
The symbols are the numerical results, and the black solid lines are from Eq~(\ref{eqn_prob_broad}) and (\ref{NarrowParam}). 
The analytical curves are only shown over their range of validity, $E_{\rm b}\leq E \leq E_{\rm eff}$.
Equation~(\ref{Eq:SimpleScaling}) is also shown for $r_{\rm eff}=-2000$~a.u. with
${\rm Im}A$ fitted to be $1.1\times 10^{-4}$~a.u. The parameters ${\rm Re}A$ are directly 
calculated from the short-range three-body potentials under the single-channel approximation, 
giving ${\rm Re}A$=50, 57, and 180~a.u. for $r_{\rm eff}=-20$, --200, and --2000~a.u., respectively.
The parameter $\alpha$ is then obtained by fitting to be $0.25$, consistent with Ref~\cite{WangPrepNarrow}.
(b) The recombination probability from numerical calculations after thermal averaging.
}
\label{fig_recomb_narrow}
\end{figure}

We calculate $K_3$ for identical bosons for $a<0$ numerically up to the short-range energy $E_{\rm s}$.
As shown in Fig.~\ref{fig_recomb_narrow}(a), the recombination probability ${\cal P}$ 
at $E\approx E_{\rm s}$ has only a 
relatively weak dependence on $r_{\mathrm{eff}}$ and $a$, 
since in our model the short-range behavior does not have strong dependence on $r_{\mathrm{eff}}$ or $a$. 
From our numerical calculations, we deduce that the energy dependence changes to ${\cal P}\propto k$ in the energy range $E_{\mathrm{eff}}< E< E_{\rm s}$. 
Therefore as the energy gets smaller, the recombination probability decreases monotonically
until $E=E_{\mathrm{eff}}$, where oscillatory behavior similar to Eq.~(\ref{eqn_prob_broad}) takes over. 
Compared to a broad resonance ($r_{\rm eff}$=$-20$~a.u. in the figure), %BDE
${\cal P}$ is suppressed by a factor of $1/|r_{\mathrm{eff}}|$ around $E\approx E_{\mathrm{eff}}$ before it 
connects to the oscillatory behavior at lower energy. 
%YW BDE
Considering that $K_3$ at a fixed $a$ has the same energy dependence for both broad and narrow Feshbach resonances
for $E< E_{\mathrm{eff}}$,
the recombination rates at lower energies are all suppressed by $1/|r_{\mathrm{eff}}|$.

To understand this observed $1/|r_{\rm eff}|$ suppression,  
we follow the analytical procedure introduced in Ref.~\cite{WangPrepNarrow}. 
The Coulomb-like potential in the region $|r_{\mathrm{eff}}|\ll R\ll |a|$ takes the form
\begin{eqnarray}
W_{\nu\nu}\approx \frac{c_0}{2\mu |r_{\rm eff}|R},
\end{eqnarray}
with $c_0$ a non-universal constant on the order of unity that can be positive or negative~\cite{WangPrepNarrow}.  We found
in \cite{WangPrepNarrow} that simply setting $W_{\nu\nu}$ to zero gave analytic results consistent with our numerical results
and is justified by the fact that $W_{\nu\nu}$ is not universal, changing from attractive to repulsive, while the numerical rates are universal.
To derive analytic expressions for the rates that connect most closely with those for
a broad resonance, we write the wave function in the region $r_0 \ll R \ll |r_{\mathrm{eff}}|$ as 
\begin{eqnarray}
F_{\nu}= C\sin(kR+\delta_{\rm eff}). 
\end{eqnarray}
For $E\ll E_{\rm s}$, we can use a low-energy expansion for $\delta_{\rm eff}$,
\begin{eqnarray}
\delta_{\rm eff}\approx-A k, 
\label{Eq:ReffScaling}
\end{eqnarray}
where we have introduced the complex three-body short-range scattering length $A$ to parametrize the 
physics at $R\leq r_0$~\cite{WangPrepNarrow}.
The recombination probability $\cal P$ can thus be derived in the same fashion 
as the broad Feshbach resonance case.
In the energy range $E_{\rm b}< E< E_{\rm eff}$, this gives the same results as Eq.~(\ref{eqn_prob_broad}),
with $r_0$ replaced by $\alpha|r_{\mathrm{eff}}|$ and the parameters $\Phi$ and $\eta$ connected to 
the real and imaginary parts of $A$:
\begin{eqnarray} 
\tan\Phi=2s_0\frac{\alpha-\mathrm{Re}A/|r_{\mathrm{eff}}|}{\alpha+\mathrm{Re}A/|r_{\mathrm{eff}}|}, \qquad
\sinh\eta=2\left|\frac{\mathrm{Im}A}{r_{\mathrm{eff}}}\right|\sin2\Phi.
\label{NarrowParam}
\end{eqnarray}
These analytical results are shown together with the numerical data in Fig.~\ref{fig_recomb_narrow}(a).
The deviation between the numerical and analytical results for $r_{\rm eff}$=$-20$ and $-200$~a.u. is 
because the universal requirement $|r_{\rm eff}|\gg r_0$ is not well satisfied.

For $E_{\mathrm{eff}}<E<E_s$, the recombination probability ${\cal P}$ has a simple scaling behavior 
independent of the scattering length:
\begin{eqnarray}
{\cal P}=4k|\mathrm{Im}A|,
\label{Eq:SimpleScaling}
\end{eqnarray}
in agreement with the numerical results.

To check the robustness of these oscillations for a thermal distribution,
we show the thermally averaged ${\cal P}$ with large $|r_{\mathrm{eff}}|$ in Fig.~\ref{fig_recomb_narrow}(b). 
It can be seen that although the modulation of the oscillations at 
low temperature is reduced, the ${\cal P}\propto k$ scaling for $E_{\mathrm{eff}}<E<E_s$ is essentially unchanged.

\subsection{Vibrational relaxation at finite energies}

\begin{figure}
\begin{center}
\includegraphics[clip=true,scale=0.9]{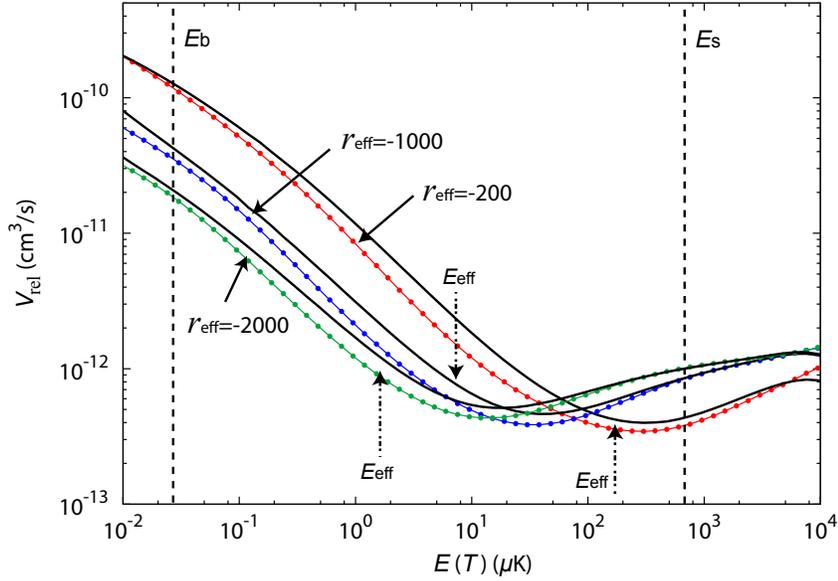}
\end{center}
\caption{The three-body relaxation rate $V_{\mathrm{rel}}$ for identical bosons near 
narrow Feshbach resonances with $a$=$10^4$~a.u. and $r_{\mathrm{eff}}$=--200, --1000. and --2000~a.u.
The rates are shown for collision energies above the three-body break-up threshold, where no 
resonant peaks are present so that the energy dependence is more easily seen.
The symbols are the numerical results, and the back solid lines are the thermally averaged results.
}
\label{fig_relax_narrow}
\end{figure}

The behavior of vibrational relaxation for identical bosons with $a>0$ changes in the same energy range 
as for three-body recombination: $E_{\rm eff}\leq E\leq E_{\rm s}$. 
Figure~\ref{fig_relax_narrow} shows that as $|r_{\mathrm{eff}}|$ increases,
instead of the $k^{-1}$ scaling seen in Eq.~(\ref{Eq:BroadVrel}) for broad Feshbach resonances, a plateau region appears in $V_{\mathrm{rel}}$, 
extending from $E_{\rm s}$ down to $E_{\rm eff}$. This change also
connects to the $1/|r_{\mathrm{eff}}|$ suppression of the rates near zero energy and can be used for observing the effective-range effect near narrow
Feshbach resonances.  Further, this plateau behavior can be derived in the same manner as Eq.~(\ref{Eq:SimpleScaling}).  

%YW BDE
Since the value of the plateau has only a weak dependence on $r_{\rm eff}$ and 
extends down to $E_{\rm eff}$, the larger 
$|r_{\rm eff}|$ is, the broader the plateau will be.
And, because $V_{\rm rel}$ for different $r_{\rm eff}$ has similar values for $E\approx E_s$, 
the fact that the plateau extends down to $E_{\rm eff}$ 
will make $V_{\mathrm{rel}}$ smaller than for a broad resonance
by a factor of $r_0/|r_{\rm eff}|$ for all energies $E< E_{\rm eff}$. 
In other words, we expect $C$ in Eq.~(\ref{Eq:BroadVrel}) to be proportional to 1/$|r_{\rm eff}|$.
The experimental observation of a plateau region in the energy dependence 
of $V_{\rm rel}$ will show the universal physics pertaining to a narrow Feshbach resonance. 
Moreover, the energy where $V_{\rm rel}$ changes from plateau to $k^{-1}$ scaling can roughly give the value 
of $|r_{\rm eff}|$.

In Fig.~\ref{fig_relax_narrow} we also show the thermally averaged relaxation rates. 
It can be seen that the plateaus in $E_{\mathrm{eff}}\leq E\leq E_{\rm s}$ 
are not dramatically changed by a thermal distribution. 

\section{Higher partial wave contributions}

To determine whether the energy-dependent features we have described will be observable, the contributions of higher partial
waves must be considered as they can easily obscure features away from threshold.
Key to this consideration is the fact that three-body recombination ($a<0$) and relaxation ($a>0$) 
are dominated by transitions at $R\leq r_0$. 
For the three-body systems where the Efimov effect does not occur for $J>0$ --- for instance, identical bosons
and $BBX$ systems with $\gamma\leq 38.61$~\cite{MassDIncao2006} ---
the effective hyperspherical potentials are all repulsive for $R>r_0$, leading
to substantial suppression of the inelastic rates when $E \ll E_{\rm s}$. 
Following the same method used for deriving the energy dependence above, 
we find that the recombination probabilities for $J>0$ scale like $(k r_0)^{2p_0}$, 
where $p_0$ is a universal constant that increases with $J$~\cite{MassDIncao2006}.  
For identical bosons, $p_0$ is always larger than $2$. 
For $E\ll E_{\rm s}$ ($kr_0\ll 1$), ${\cal P} {(J>0)}$ are all negligible. 
Moreover, we have confirmed 
this conclusion for a few cases numerically.
For vibrational relaxation, there is similar suppression for $J>0$ contributions.

\section{Summary}

We have studied inelastic three-body collisions and have found that they behave universally
for collision energies well out of the ultracold regime --- up to nearly a milli-Kelvin in the examples shown.
Moreover, we have identified much of the universal energy dependence as a manifestation of the Efimov effect,
opening up new dimensions on Efimov physics.  Significantly, we have been able to derive analytic expressions,
which were motivated and verified by numerical calculations, for these rates in terms of a few parameters.
And, we have done so for both broad and narrow resonances.  Our study thus brings universal three-body physics 
out of the ultracold regime, suggesting even
richer universal physics can be discovered both theoretically and experimentally.

\section*{Acknowledgments}
We thank J.P. D'Incao for early discussions of this problem. %BDE
This work was supported in part by the National Science Foundation and in part by Air Force Office of Scientific Research.
Y. W. also acknowledge the support from the National Science Foundation under Grant No. PHY0970114.

\section*{References}

\bibliography{References}{}

\begin{thebibliography}{10}
\bibitem{Unit}
Collision energies are reported in Kelvin using the conversion {$T = E/k_b$}.
\end{thebibliography}
\bibliographystyle{unsrt}

\end{document}